\documentclass[12pt]{article}
\usepackage{amsmath}
\usepackage{amsfonts}
\usepackage{amssymb}
\usepackage{latexsym}
\usepackage{epsfig}
\usepackage{graphicx}
\usepackage{oldgerm}
\usepackage{theorem}
\usepackage{bm}

\def\C{\mathbb{C}}
\def\R{\mathbb{R}}
\def\N{\mathbb{N}}
\def\B{\mathbb{B}}
\def\Z{\mathbb{Z}}
\def\D{\mathbb{D}}
\def\HH{\mathbb{H}}
\def\bE{\mathbf{E}}
\def\bP{\mathbf{P}}

\def\cB{\mathcal{B}}
\def\cI{\mathcal{I}}
\def\cD{\mathcal{D}}
\def\cF{\mathcal{F}}
\def\cU{\mathcal{U}}

\def\sF{\mathsf{F}}
\def\sH{\mathsf{H}}
\def\sh{\mathsf{h}}
\def\sB{\mathsf{B}}

\def\Conf{\mathrm{Conf}}
\def\var{\mathrm{var}}
\def\vol{\mathrm{vol}}
\def\Re{\mathrm{Re}\,}
\def\Im{\mathrm{Im}\,}
\def\rR{\mathrm{R}}
\def\rI{\mathrm{I}}
\def\rN{\mathrm{N}}

\def\rO{\mathrm{O}}
\def\bra{\langle}
\def\ket{\rangle}
\def\dis={\stackrel{\rm d}{=}}

\newtheorem{thm}{Theorem}[section]
\newtheorem{lem}[thm]{Lemma}
\newtheorem{cor}[thm]{Corollary}
\newtheorem{prop}[thm]{Proposition}

\newtheorem{df}[thm]{Definition}

\newcommand{\SSC}[1]{\section{#1}\setcounter{equation}{0}}
\newcommand{\qed}{\hbox{\rule[-2pt]{3pt}{6pt}}}

\begin{document}

\title{
Local number variances and 
hyperuniformity 
of the Heisenberg family of 
determinantal point processes
}
\author{
Takato Matsui
\footnote{
Department of Physics,
Faculty of Science and Engineering,
Chuo University, 
Kasuga, Bunkyo-ku, Tokyo 112-8551, Japan;
e-mail: matsui@phys.chuo-u.ac.jp
}
\quad
Makoto Katori
\footnote{
Department of Physics,
Faculty of Science and Engineering,
Chuo University, 
Kasuga, Bunkyo-ku, Tokyo 112-8551, Japan;
e-mail: katori@phys.chuo-u.ac.jp
} 
\quad
Tomoyuki Shirai 
\footnote{
Institute of Mathematics for Industry, 
Kyushu University, 
744 Motooka, Nishi-ku,
Fukuoka 819-0395, Japan; 
e-mail: shirai@imi.kyushu-u.ac.jp
}
}
%%%%%%%%%%%%%%%%%%%%%%%%%%%%%%%%
\date{9 March 2021}
%%%%%%%%%%%%%%%%%%%%%%%%%%%%%%%
\pagestyle{plain}
\maketitle

\begin{abstract}
The bulk scaling limit of eigenvalue distribution 
on the complex plane ${\mathbb{C}}$
of the complex Ginibre random matrices provides
a determinantal point process (DPP).
This point process is a typical example of 
disordered hyperuniform system characterized
by an anomalous suppression of large-scale 
density fluctuations.
As extensions of the Ginibre DPP, we consider
a family of DPPs defined on the $D$-dimensional 
complex spaces ${\mathbb{C}}$, $D \in {\mathbb{N}}$, in which
the Ginibre DPP is realized when $D=1$. 
This one-parameter family ($D \in {\mathbb{N}}$) of DPPs 
is called the Heisenberg family,
since the correlation kernels are
identified with the Szeg\H{o} kernels for the
reduced Heisenberg group.
For each $D$, using the modified Bessel functions, 
an exact and useful expression
is shown for the local number variance
of points included in a ball with radius $R$
in ${\mathbb{R}}^{2D} \simeq {\mathbb{C}}^D$. 
We prove that any DPP in the Heisenberg family is in the
hyperuniform state of Class I, 
in the sense that the number variance 
behaves as $R^{2D-1}$ as $R \to \infty$.
Our exact results provide asymptotic expansions 
of the number variances in large $R$.

\vskip 0.5cm


\end{abstract}
%%%%%%%%%%%%%%%%%%%%%%%%%%%%%%
%\newpage
%%%%%%%%%%%%%%%%%%%%%%%
%\footnotesize 
%\tableofcontents
%%%%%%%%%%%%%%%%%%%%%%%
\vspace{3mm}
\normalsize
%\setlength{\baselineskip}{1cm}

%%%%%%%%%%%%%%%%%%%%%%%%%%%%%%%%%%%%%%%%%%%%%%%%%%%%%%%%%%
%%%  SEC1 %%%%%%%%%%%%%%%%%%%%%%%%%%%%%%%%%%%%%%%%%%%
%%%%%%%%%%%%%%%%%%%%%%%%%%%%%%%%%%%%%%%%%%%%%%%%%%%%%%%%%%
\SSC
{Introduction and Main Results} \label{sec:Introduction}%%%%%%%%
%%%%%%%%%%%%%%%%%%%%%%%%%%%%%%%%%%%%%%%%%%%%%%%%%%

We consider the $d$-dimensional Euclid space
$\R^d$, $d \in \N:=\{1,2, \dots\}$, 
or the $D$-dimensional complex space $\C^D$, $D \in \N$
as a base space $S$. 
We assume that $S$ is associated with a reference measure
$\lambda$.
We consider an \textit{infinite point process} on $S$,
which is expressed by an infinite sum of delta measures 
concentrated on a set of random points
$X_i, i \in \N$,
\begin{equation}
\Xi = \sum_{i: i \in \N} \delta_{X_i}.
\label{eqn:Xi1}
\end{equation}
Here a delta measure
$\delta_{X}(\{x\})$, $x \in S$, gives $1$ if $x=X$, and 0 otherwise.
Hence the number of points included in a domain $\Lambda \subset S$
is given by
$\Xi(\Lambda):= \int_{\Lambda} \Xi(dx)=\sum_{i: X_i \in \Lambda} 1$.
We assume that for any 
bounded domain $\Lambda \subset S$,
$\Xi(\Lambda) < \infty$; 
that is, accumulation of points does not occur, and 
with respect to the reference measure $\lambda(dx)$
the point process has a finite density
$\rho_1(x) < \infty$ 
at almost every $x \in S$.
We consider a \textit{homogeneous} point process
in the sense that $\rho_1(x) \lambda(dx)= {\rm const.} \times dx$,
$x \in S$, where $dx$ denotes the Lebesgue measure on $S$.
The above assumption implies that 
for a bounded domain $\Lambda \subset S$
the expectation of 
$\Xi(\Lambda)$ is proportional to
the volume $\vol(\Lambda)$ of $\Lambda$; 
$\bE[\Xi(\Lambda)] \propto \vol(\Lambda)$.
Now we consider the number variance 
in the domain $\Lambda$,
\[
\var[\Xi(\Lambda)] := \bE[(\Xi(\Lambda)- \bE[\Xi(\Lambda)])^2],
\]
which represents local density fluctuation of 
point process $\Xi$. 
If the points are non-correlated and
given by a Poisson process, then 
$\var[\Xi(\Lambda)] \propto \vol(\Lambda)$.

Recently in condensed matter physics and related material sciences,
correlated particle systems are said to be in a
\textit{hyperuniform state} when density fluctuations
are anomalously suppressed in large-scale limit.
The bounded domain $\Lambda$ is regarded as
a observation \textit{window} to measure 
density fluctuation of the system.
For an infinite random point process $\Xi$, 
the \textit{hyperuniformity} is defined by 
\begin{equation}
\lim_{\Lambda \to S}
\frac{\var[\Xi(\Lambda)]}{\bE[\Xi(\Lambda)]}
=0. 
\label{eqn:HU}
\end{equation}
This means that 
the number variance of points grows
more slowly than the window volume in the limit
such that the window covers whole of the space
$\Lambda \to S$. 
See \cite{GL17,Tor18} and references therein. 
Moreover, 
Torquato \cite{Tor18} proposed
the three hyperuniformity classes for point processes
concerning asymptotics of number variances.
In order to clearly state this classification,
here we assume that $S=\R^d$ and
$\Lambda=\B^{(d)}_R$, $d \in \N$, where 
$\B^{(d)}_R$ denotes a ball in $\R^d$ with radius $R >0$ 
centered at the origin;
$\B^{(d)}_R :=\{x \in \R^d : |x| < R\}$. 
The volume of the ball is given by
\begin{equation}
\vol(\B^{(d)}_R) = \frac{\pi^{d/2}}{\Gamma(d/2+1)} R^d, 
\label{eqn:vol}
\end{equation}
where the gamma function is defined by
$\Gamma(z) :=\int_0^{\infty} e^{-u} u^{z-1} du$, 
$\Re z >0$ and it satisfies the functional equation
$\Gamma(z+1)=z \Gamma(z)$ 
with $\Gamma(1)=1$ and $\Gamma(1/2)=\sqrt{\pi}$. 
We consider a series of balls with increasing $R$,
$\{\B^{(d)}_R \}_{R >0}$, and
the hyperuniform states are classified 
as follows; 
\begin{align*}
&\mbox{Class I} : \qquad \, \, \,
\var[\Xi(\B^{(d)}_R)]
\asymp R^{d-1},
\nonumber\\
&\mbox{Class II} : \qquad \, \,
\var[\Xi(\B^{(d)}_R)]
\asymp R^{d-1} \log R,
\nonumber\\
&\mbox{Class III} : \qquad
\var[\Xi(\B^{(d)}_R)]
\asymp R^{d-\alpha}, \quad 0<\alpha< 1,
\quad \mbox{as $R \to \infty$}.
\end{align*}
Here $f(R) \asymp g(R)$ means that there
are finite positive constants $c_1$ and $c_2$ such that
$c_1 g(R) < f(R) < c_2 g(R)$.
The above characterization of these classes
will be similarly described for any 
series of windows $\{\Lambda_R \}_{R >0}$
labeled by a linear scale $R$ of window.

\textit{Determinantal point processes} (DPPs)
\cite{Sos00,ST03a,ST03b,HKPV06,HKPV09} 
studied in random matrix theory (RMT)
\cite{Meh04,For10}
provide a variety of examples of hyperuniform systems.
In general a DPP is specified by a triplet 
$(\Xi, K, \lambda(dx))$ \cite{KS19+}, 
where $\Xi$ is a random measure
(\ref{eqn:Xi1}) representing a point process, 
$K$ is a continuous function
$S \times S \to \C$ called the
\textit{correlation kernel},
and $\lambda(dx)$ is a reference measure defined on $S$. 
In Section \ref{sec:DPP} below a precise definition of DPP will be given.

The most studied DPP in RMT may be the
\textit{sinc} (\textit{sine}) \textit{determinantal point process} (DPP), 
$(\Xi_{\rm sinc}, K_{\rm sinc}, dx)$ on $S=\R$,
where $K_{\rm sinc}(x,y)=\sin(x-y)/\{\pi(x-y)\}$, $x, y \in \R$. 
This DPP is obtained as the bulk scaling limit
of the eigenvalue distribution of Hermitian random matrices
in the Gaussian unitary ensemble.
It is known as a classical result in RMT that
\[
\var[\Xi_{\rm sinc}(\B^{(1)}_R)] 
\sim 
\frac{\log R}{\pi^2} 
\quad \mbox{as $R \to \infty$}.
\]
See, for instance, 
\cite[Section 16.1]{Meh04},
\cite{CL95},
\cite{Sos00b}
\cite{Sos02}
\cite[Remark 5.8]{ST03a}.
In the present paper $f(R) \sim g(R)$ as $R \to \infty$ means
$\lim_{R \to \infty} f(R)/g(R)=1$. 
That is, the sinc DPP is in Class II of hyperuniformity.
Torquato et al.\cite{TSZ08} and Torquato \cite{Tor18}
studied one-parameter ($d \in \N$) family of DPPs
called the \textit{Fermi-sphere point processes},
which gives the sinc DPP when $d=1$.
They proved that the Fermi-sphere point processes
are in Class II for general $d \in \N$.

An example of infinite DPP in Class I
of hyperuniformity is also provided in RMT.
It is the DPP on $\C$ called the
\textit{Ginibre DPP}, 
$(\Xi_{\rm Ginibre}, K_{\rm Ginibre}, \lambda_{\rN(0,1; \C)}(dx))$, 
which is obtained as the
bulk scaling limit of eigenvalue distribution of 
non-Hermitian random matrices in the
complex Ginibre ensemble \cite{Gin65}. 
Here the correlation kernel is
given by $K_{\rm Ginibre}(x,y)=e^{x \overline{y}}$, $x, y \in \C$,
where $\overline{y}$ denotes the complex conjugate of $x$,
and $\lambda_{\rN(0,1; \C)}(dx)$ is the
\textit{complex} standard normal distribution; 
$\lambda_{\rN(0,1;\C)}(dx)=e^{-|x|^2} dx/\pi$.
A disk on $\C$ centered at the origin with radius $R$,
$\D_R :=\{x \in \C: |x| < R \}$, is identified with 
$\B^{(2)}_R \subset \R^2$.
One of the present authors proved \cite{Shi06}
\[
\var[\Xi_{\rm Ginibre}(\B^{(2)}_R) ]
\sim \frac{R}{\sqrt{\pi}} \quad
\mbox{as $R \to \infty$}.
\]
See also \cite{OS08,Shi15,Tor18}. 
In the present paper we will report extensions of 
this result in $S=\C$ to DPPs in the
higher-dimensional complex spaces
$S=\C^D, D=2,3 \dots$.

When $S=\C^D, D \in \N$, each coordinate
$x \in \C^D$ has $D$ complex components;
$x=(x^{(1)}, \dots, x^{(D)})$ with
$x^{(\ell)}=\Re x^{(\ell)}+ \sqrt{-1} \Im x^{(\ell)}$, $\ell=1, \dots, D$.
In order to clearly describe such a complex structure, 
we set $x_{\rR}:=(\Re x^{(1)}, \dots, \Re x^{(D)})$,
$x_{\rI}:=(\Im x^{(1)}, \dots, \Im x^{(D)}) \in \R^D$,
and write $x=x_{\rR}+ \sqrt{-1} x_{\rI}$ in this paper.
The Lebesgue measure on $\C^D$ is given by
$d x = d x_{\rR} d x_{\rI} 
:= \prod_{\ell=1}^D d \Re x^{(\ell)} d \Im x^{(\ell)}$.
For $x = x_{\rR}+ \sqrt{-1} x_{\rI}$,
$y=y_{\rR}+ \sqrt{-1} y_{\rI} \in \C^D$,
we use the \textit{standard Hermitian inner product}; 
\[
x \cdot \overline{y} 
:= (x_{\rR}+ \sqrt{-1} x_{\rI}) \cdot (y_{\rR}-\sqrt{-1} y_{\rI})
=(x_{\rR} \cdot y_{\rR}+x_{\rI} \cdot y_{\rI})
-\sqrt{-1} (x_{\rR} \cdot y_{\rI}- x_{\rI} \cdot y_{\rR}).
\]
Notice that if $x=x_{\rR}, y=y_{\rR} \in \R^D$,
then 
$x \cdot \overline{y}=x_{\rR} \cdot y_{\rR} 
:=\sum_{\ell=1}^D \Re x^{(\ell)} \Re y^{(\ell)}$. 
We define the norm by
$|x| := \sqrt{x \cdot \overline{x}} = \sqrt{|x_{\rR}|^2+|x_{\rI}|^2}$,
$x \in \C^D$.
Hence the $D$-dimensional disk 
$\{x \in \C^D: |x| < R\}$ 
centered at the origin with radius $R$ on $\C^D$ will be
identified with $\B^{(d)}_{R}$ in $\R^d$ with $d=2D, D \in \N$.
On $\C^D$ the reference measure is given by
the $D$-dimensional extension of $\lambda_{\rN(0,1; \C^D)}(dx)$, 
\begin{align}
\lambda_{\rN(0, 1; \C^D)}(d x)
&:= \prod_{i=1}^D \lambda_{\rN(0, 1; \C)}(d x^{(i)})
\nonumber\\
&= \frac{e^{-|x|^2}}{\pi^D} dx 
= \frac{e^{-(|x_{\rR}|^2+|x_{\rI}|^2)}}{\pi^D} dx_{\rR} dx_{\rI}. 
\label{eqn:lambdaN}
\end{align}

The one-parameter family ($D \in \N$) of DPPs
studied in this paper is the
\textit{Heisenberg family of DPPs} 
defined on $\C^D$ as follows.

%%%%%%%%%%%%%%%%%%%%
\begin{df}
\label{thm:HeisenbergDPP}
The Heisenberg family of DPPs 
is defined by
$(\Xi_{\sH_D}, K_{\sH_D}, \lambda_{\rN(0, 1; \C^D)})$
on $\C^D$, $D \in \N$
with the correlation kernel 
\begin{equation}
K_{\sH_D}(x, y)
=e^{x \cdot \overline{y}},
\quad x, y \in \C^D.
\label{eqn:KHD}
\end{equation}
\end{df}
\vskip 0.3cm
%%%%%%%%%%%%%%%%%%%%%

Note that $K_{\sH_D}$ is hermitian;
$\overline{K_{\sH_D}(x,y)}=K_{\sH_D}(y,x)$, $x, y \in \C^D$. 
The kernels in this form on $\C^D, D \in \N$ have been
studied by Zelditch and his coworkers
(see \cite{Zel01,BSZ00} and references therein),
who identified them with the Szeg\H{o} kernels
for the \textit{reduced Heisenberg group} $\sH_D^{\rm red}$.
This is the reason why we call the DPPs associated with
(\ref{eqn:KHD}) 
the \textit{Heisenberg family of DPPs} on $\C^D, D \in \N$
\cite{KS19+}.
This family includes the complex Ginibre DPP 
\cite{Gin65,HKPV06,OS08,HKPV09,Shi15,Tor18,Kat19b} as the
lowest dimensional case with $D=1$.
A brief review of the representation theory of
the Heisenberg group 
$\sH_D$ is given in 
Appendix \ref{sec:HeisenbergGroup}.
There the Bargmann--Fock representation
of $\sH_D$ is explained and 
the correlation kernel (\ref{eqn:KHD}) is
realized as the \textit{reproducing kernel} of
the Bargmann--Fock space $\cF_D$.
It should be noted that \cite{KS19+}
if we follow a similar reasoning, the Fermi-sphere 
point processes studied by 
Torquato et al. \cite{TSZ08,Tor18}
can be called the 
\textit{Euclidean family of DPPs},
since the correlation kernels in these DPPs can be regarded
as the Szeg\H{o} kernels
for the reduced Euclidean motion group 
\cite{Zel01,SZ02,Zel09,CH15}. 

Define the modified Bessel function of the first kind 
\cite{Wat44,NIST10} by
\begin{equation}
I_{\nu}(z):=\left( \frac{z}{2} \right)^{\nu}
\sum_{n=0}^{\infty} \frac{(z/2)^{2n}}{n! \Gamma(\nu+n+1)},
\quad z \in \C \setminus (-\infty, 0]. 
\label{eqn:mBessel}
\end{equation}
We prove the following.
%%%%%%%%%%%%%%%%%%%%%%%%%
\begin{prop}
\label{thm:main1}
For the Heisenberg family of DPPs,
$(\Xi_{\sH_D}, K_{\sH_D}, \lambda_{\rN(0, 1; \C^D)})$ 
on $\C^D, D \in \N$, 
\begin{align}
\var[\Xi_{\sH_D}(\B^{(2D)}_R)]
&=\frac{R^{2D} e^{-2 R^2}}{D!} 
\sum_{n=0}^{D-1} 
[I_n(2R^2) +I_{n+1}(2R^2)] 
\nonumber\\
&=\frac{R^{2D} e^{-2 R^2}}{D!} 
\left[ I_0(2R^2) + 2 \sum_{n=1}^{D-1} I_n(2R^2) +I_D(2R^2)
\right], \quad R >0.
\label{eqn:main1}
\end{align}
\end{prop}
%%%%%%%%%%%%%%%%%%%%%%%

\vskip 0.3cm
%%%%%%%%%%%%%%%%%
\noindent{\bf Remark 1} \,
When $D=1$, (\ref{eqn:main1}) gives 
\[
\var[\Xi_{\sH_1}(\B^{(2)}_R)]
=\var[\Xi_{\rm Ginibre}(\B^{(2)}_R)]
=R^{2} e^{-2 R^2}
[I_0(2R^2) +I_1(2R^2)], 
\]
which 
is identified with
Eq.(249) in \cite{Tor18} calculated for
the complex Ginibre DPP.
%%%%%%%%%%%%%%%%%%%%%%%%%%%%

\vskip 0.3cm
%%%%%%%%%%%%%%%%%
\noindent{\bf Remark 2} \,
If we use the following type of hypergeometric function
\begin{equation}
{_{2} F_2}(a_1, a_2; b_1, b_2; x)
:=\sum_{n=0}^{\infty} \frac{(a_1)_n (a_2)_n}{(b_1)_n (b_2)_n} 
\frac{x^n}{n!}
\label{eqn:HG1}
\end{equation}
with the Pochhammer symbols,
$(a)_0 :=1$,
$(a)_n :=a(a+1) \cdots (a+n-1), n \in \N$, 
the number variances (\ref{eqn:main1}) are expressed as
\begin{align}
&\var[\Xi_{\sH_D}(\B^{(2D)}_R)]
\nonumber\\
& \qquad 
=\frac{R^{2D}}{\Gamma(D+1)} 
\left[ 1 - \frac{R^{2D}}{\Gamma(D+1)}
{_2 F_2} (
D, D+1/2; D+1, 2D+1; -4 R^2 ) \right],
\label{eqn:HG2}
\end{align}
$D \in \N, R > 0$. 
We found that the expressions (\ref{eqn:main1}) 
using the modified Bessel functions
with argument $2R^2$ are more useful than (\ref{eqn:HG2}) to 
derive the following results.
%%%%%%%%%%%%%%%%%%%%%%%%%%%%
\vskip 0.3cm

%%%%%%%%%%%%%%%%%%%%%
\begin{thm}
\label{thm:main2}
Any DPP in the Heisenberg family, 
$(\Xi_{\sH_D}, K_{\sH_D}, \lambda_{\rN(0, 1; \C^D)})$
on $\C^D$, $D \in \N$, 
is in the hyperuniform state of Class I
such that
\begin{equation}
\lim_{R \to \infty} R \frac{\var[\Xi_{\sH_D}(\B^{(2D)}_R)]}
{\bE[\Xi_{\sH_D}(\B^{(2D)}_R)]}
=\frac{D}{\sqrt{\pi}}.
\label{eqn:C0}
\end{equation}
Moreover, for each $D \in \N$, 
the following asymptotic expansion holds, 
\begin{equation}
\frac{\var[\Xi_{\sH_D}(\B^{(2D)}_R)]}
{\bE[\Xi_{\sH_D}(\B^{(2D)}_R)]}
\sim
\frac{D}{\sqrt{\pi}} R^{-1}
\sum_{k=0}^{\infty} (-1)^k \frac{\alpha_k(D)}{(2k+1) k! 2^{4k}} R^{-2k}
\quad \mbox{as $R \to \infty$}, 
\label{eqn:asym}
\end{equation}
where 
\begin{equation}
\alpha_k(D) = 
\begin{cases}
1, & \quad \mbox{if $k=0$},
\cr
\displaystyle{
\prod_{\ell=1}^k \{4 D^2-(2\ell-1)^2\}
=\prod_{\ell=-k+1}^k(2D+2 \ell-1)
},
& \quad \mbox{if $k \in \N$}. 
\end{cases}
\label{eqn:alpha1}
\end{equation}
\end{thm}
%%%%%%%%%%%%%%%%%%%%%%%%%%%%

\vskip 0.3cm
%%%%%%%%%%%%%%%%%
\noindent
{\bf Remark 3} \,
The Heisenberg family of DPPs belongs to
a wider class of DPPs called
the infinite \textit{Weyl--Heisenberg ensemble}
studied by Abreu et al. \cite{AGR16,APRT17,AGR19}.
In the present setting and notations,  
a DPP in the Weyl--Heisenberg ensemble is expressed by
$(\Xi_{\rm WH}, K_{\rm WH}^g, dx)$ 
on $S=\C^D \simeq \R^{2D}$, $D \in \N$, with the
correlation kernel in the form,
\[
K_{\rm WH}^g(x, y) 
= \int_{\R^D} g(u-x_{\rR}) \overline{g(u-y_{\rR})}
e^{2 \sqrt{-1} (x_{\rI}-y_{\rI}) \cdot u} du,
\quad x, y \in \C^D,
\]
where a function $g$ on $\R^D$ satisfies some conditions
\cite{AGR16,APRT17,AGR19}. 
See also Section 2.6 of \cite{KS19+}. 
We can verify that if $g$ is chosen as
\begin{equation}
G(\zeta)=\left(\frac{2}{\pi} \right)^{D/4}
\frac{e^{-\zeta^2}}{\pi^{D/2}},
\quad \zeta \in \R^D,
\label{eqn:G1}
\end{equation}
then the conditions are satisfied and
the following equality is obtained, 
\[
K_{\rm WH}^G(x, y)
=\frac{e^{\sqrt{-1} x_{\rR} \cdot x_{\rI}}}
{e^{\sqrt{-1} y_{\rR} \cdot y_{\rI}}}
\sqrt{ \frac{ e^{-|x|^2} }{\pi^D}} 
K_{\sH_D}(x, y) 
\sqrt{ \frac{ e^{-|y|^2} }{\pi^D}}, 
\quad x, y \in \C^D.
\]
The factor 
$e^{\sqrt{-1} x_{\rR} \cdot x_{\rI}}/
e^{\sqrt{-1} y_{\rR} \cdot y_{\rI}}$
is irrelevant for DPP and
this equality proves the equivalence
between $(\Xi_{\sH_D}, K_{\sH_D}, \lambda_{\rN(0, 1; \C^D)})$
and
$(\Xi_{\rm WH}, K_{\rm WH}^G, dx)$
on $\C^D$, $D \in \N$. 
It was proved in \cite[Theorem 5.8]{APRT17} that
any DPP in the Weyl-Heisenberg ensemble is
in the hyperuniform state of Class I.
Notice that we have determined the coefficient
$D/\sqrt{\pi}$ of the dominant term in $R \to \infty$
as (\ref{eqn:C0}) and
completed the asymptotic expansion 
(\ref{eqn:asym}) with (\ref{eqn:alpha1})
for the Heisenberg family of DPPs 
in the above theorem.

\vskip 0.3cm
%%%%%%%%%%%%%%%%%
\noindent{\bf Remark 4} \,
Since the present Heisenberg family of point processes 
is determinantal, higher cummulants of 
$\Xi_{\sH_D}(\B^{(2D)}_R)$ can be directly calculated 
for finite values of $R$ and their asymptotics in $R \to \infty$
will be evaluated.
See \cite{CL95,For99,Sos00,Sos02,ST03a,Shi06,OS08}
for general formulas of cumulants of linear statistics
and their generating functions
as well as applications to the sinc DPP
and the Ginibre DPP (\textit{i.e.}, the Heisenberg DPP
with $D=1$).
In the present paper we concentrated on 
expectations and variances, since 
hyperuniformity (\ref{eqn:HU}) defined 
by these first two cumulants is focused.
By Proposition 2.4 in \cite{Shi06} 
(see also Theorem 24 in \cite{HKPV06}),
the divergence 
$\var[\Xi_{\sH_D}(\B^{(2D)}_R)] \to \infty$
as $R \to \infty$ implies that
$\Xi_{\sH_D}(\B^{(2D)}_R)/
\bE[\Xi_{\sH_D}(\B^{(2D)}_R)] \to 1$ 
almost surely and the
\textit{central limit theorem} holds; as $R \to \infty$,  
$(\Xi_{\sH_D}(\B^{(2D)}_R)-\bE[\Xi_{\sH_D}(\B^{(2D)}_R)])/
\var[\Xi_{\sH_D}(\B^{(2D)}_R)]$
converges in distribution to the standard normal
distribution $\rN(0,1)$.
%%%%%%%%%%%%%%%%%%%
\vskip 0.3cm

%%%%%%%%%%%%%%%%%
\noindent{\bf Remark 5} \,
Applying the \textit{duality relation between DPPs}
(see Theorem 2.6 in \cite{KS19+}), 
we can evaluate 
$\var[\Xi_{\sH_D}(\Lambda)]/\bE[\Xi_{\sH_D}(\Lambda)]$
for windows 
which are different from balls.
A \textit{polydisk} $\Delta^{(D)}_R$ of radius $R>0$
in $\C^D$, $D \in \N$
is defined by
$\Delta^{(D)}_R := \{x=(x_1, \dots, x_D) \in \C^D :
|x_i| < R, i=1, \dots, D \}$.
We can show that
\begin{align}
& \frac{\var[\Xi_{\sH_D}(\Delta^{(D)}_R)]}
{\bE[\Xi_{\sH_D}(\Delta^{(D)}_R)]}
=1 - \left( 1- 
\frac{\var[\Xi_{\sH_1}(\B^{(2)}_R)]}
{\bE[\Xi_{\sH_1}(\B^{(2)}_R)]}
 \right)^D
\nonumber\\
&\qquad \sim
\frac{D}{\sqrt{\pi}} R^{-1}
\left[ 1
- \frac{D-1}{2\sqrt{\pi}} R^{-1}
+\frac{1}{2} \left\{
\frac{(D-1)(D-2)}{3 \pi} - \frac{1}{8} \right\} R^{-2}
+\rO(R^{-3})
\right]
\label{eqn:polyD1}
\end{align}
as $R \to \infty$. 
The leading term in $R \to \infty$ 
is exactly the same as (\ref{eqn:C0})
for balls $\B_R^{(2D)}$, 
but the correction terms
with $R^{-k}, k \geq 1$ are different from
(\ref{eqn:asym}).
%%%%%%%%%%%%%%%%%%%
\vskip 0.3cm

The paper is organized as follows.
In Section \ref{sec:preliminaries} we will give preliminaries
for linear statistics of 
infinite point processes which are translationally invariant 
in distribution \cite{Tor18}.
There the formulas of Bessel functions
which we use in this paper are also summarized. 
In Section \ref{sec:number} we show 
useful formulas for local number variances
and the definition of DPP with additional assumptions 
is given.
Then the Heisenberg family of DPPs on $\C^D \simeq \R^{2D}$,
$D \in \N$ is studied. 
The proofs of Proposition \ref{thm:main1},
the formula (\ref{eqn:HG2}) in Remark 2, 
Theorem \ref{thm:main2}, and
the formula (\ref{eqn:polyD1}) in Remark 5 are
given in Section \ref{sec:proofs}.
In Appendix \ref{sec:HeisenbergGroup}
a brief review of the representation theory of
the Heisenberg group $\sH_D$ \cite{Fol89,Ste93,Gro01} is given. 

%%%%%%%%%%%%%%%%%%

%%%%%%%%%%%%%%%%%%%%%%%%%%%%%%%%%%%%%%%%%%%%%%%%%%%%%%%%%%
%%%  SEC2 %%%%%%%%%%%%%%%%%%%%%%%%%%%%%%%%%%%%%%%%%%%
%%%%%%%%%%%%%%%%%%%%%%%%%%%%%%%%%%%%%%%%%%%%%%%%%%%%%%%%%%
\SSC
{Preliminaries} \label{sec:preliminaries}%%%%%%%%
%%%%%%%%%%%%%%%%%%%%%%%%%%%%%%%%%%%%%%%%%%%%%%%%%
\subsection{Correlation functions and variances}
\label{sec:correlation}
%%%%%%%%%%%%%%%%%%%%%%%%%%%%%%%%%%%%%%%%%%%%%%%%%

The configuration space of point process $\Xi=\Xi(\cdot)$
is given by
\[
\Conf(S)
=\Big\{ \xi = \sum_i \delta_{x_i} : \mbox{$x_i \in S$,
$\xi(\Lambda) < \infty$ for all bounded set $\Lambda \subset S$} 
\Big\}.
\]
If $\Xi(\{ x \}) \in \{0, 1\}$ for any point $x \in S$,
then the point process is said to be \textit{simple}.
Let
$\cB_{\rm c}(S)$ be the set of 
all bounded measurable 
complex functions on $S$ of compact support
and for 
$\xi \in \Conf(S)$ and $\phi \in \cB_{\rm c}(S)$
we set 
\[
\bra \xi, \phi \ket :=\int_S \phi(x) \, \xi(dx) 
=\sum_i \phi(x_i).
\]
Random variables written in this form
are generally called
\textit{linear statistics} of a point process $\Xi$
\cite[Section 16.1]{Meh04}
\cite[Definition 14.3.1]{For10}.
For a point process $\Xi$, 
if there exists a non-negative measurable function
$\rho_1$ such that
\begin{equation}
\bE[\bra \Xi, \phi \ket ]
=\int_S \phi(x) \rho_1(x) \lambda(dx)
\quad \forall \phi \in \cB_{\rm c}(S),
\label{eqn:1_correlation}
\end{equation}
$\rho_1$ is called the \textit{first correlation function} of $\Xi$ 
with respect to the reference measure $\lambda$.
By definition, $\rho_1(x)$ gives the density of point
at $x \in S$ with respect to $\lambda(dx)$.
For $n \in \N$, 
from $\xi \in \Conf(S)$ we define
$\xi_n := \sum_{i_1, \dots, i_n : i_j \not= i_k, j \not=k} 
\delta_{x_{i_1}} \cdots \delta_{x_{i_n}}$
and denote the $n$-product measure of $\lambda$ by 
$\lambda^{\otimes n}$;
$\lambda^{\otimes n}(dx_1 \cdots dx_n)
:= \prod_{i=1}^n \lambda(dx_i)$. 
For a point process $\Xi$, 
if there exists a symmetric, non-negative measurable function
$\rho_n$ on $S^n$ such that
\[
\bE[\bra \Xi_n, \phi \ket ]
=\int_{S^n} \phi(x_1, \dots, x_n) 
\rho_n(x_1, \dots, x_n) \lambda^{\otimes n} (dx_1 \cdots dx_n)
\quad \forall \phi \in \cB_{\rm c}(S^n),
\]
we say that $\rho_n$ is the 
\textit{$n$-th correlation function} of $\Xi$ with respect to $\lambda$.

We put the assumptions.
\begin{description}
\item{\bf (A1)} 
The point process $\Xi$ on $(S, \cB_{\rm c}(S), \lambda)$
has the first and the second correlation functions. 
\end{description}

The following is readily proved by
the definition of correlation functions given above
(see, for instance, \cite{TS03,TS06,OS08,GL17,Tor18}).
%%%%%%%%%%%%%%%%%%%%%%%%%%%
\begin{lem}
\label{thm:variance}
Assume {\bf (A1)}. 
For $\phi \in \cB_c(S)$, the variance 
\[
\var[\bra \Xi, \phi \ket]
:= \bE[ |\bra \Xi, \phi \ket - \bE[\bra \Xi, \phi \ket] |^2]
\]
is expressed as 
\begin{equation}
\var[\bra \Xi, \phi \ket]
= \int_S |\phi(x)|^2 \rho_1(x) \lambda(dx)
+\int_{S \times S} \phi(x) \overline{\phi(y)}
(\rho_2(x, y)-\rho_1(x) \rho_1(y)) \lambda^{\otimes 2} (dx dy).
\label{eqn:var1}
\end{equation}
\end{lem}
%%%%%%%%%%%%%%%%%%%%%%%%%%%

From now on we consider the case in which
$S = \R^d$ with $d \in \N$.
We put a further assumption.
%%%%%%%%%%%%%%%%%%%%%%%%%%%%%%%%%%%%%%%%%
\begin{description}
\item{\bf (A2)} 
The system is translationally invariant with respect to
the Lebesgue measure $dx$ on $\R^d$
in the following sense.
\begin{description}
\item{(i)} \,
The reference measure has a density $\ell(x)$ with respect to
the Lebesgue measure $dx$ on $\R^d$; 
$\lambda(dx)=\ell(x) dx, x \in \R^d$, and
\[
\rho_1(x) \ell(x) = \mbox{constant} 
=: \widetilde{\rho} \quad \forall x \in \R^d.
\]
\item{(ii)} \,
There is a measurable even function 
$g_2(x)=g_2(-x), x \in \R^d$
so that the second correlation function is written in the form
\[
\rho_2(x, y) \ell(x) \ell(y)= \widetilde{\rho}^2 g_2(x-y),
\quad x, y \in \R^d.
\]
\end{description}
\end{description}
%%%%%%%%%%%%%%%%%%%%%%%%%%%%%%%%%%%%%
The two-point function $g(x)$ is called
the \textit{unfolded 2-correlation function} \cite{For10}.
We define the following function
which is called the \textit{total correlation function}
\cite{Tor18}, 
\begin{equation}
C(x)=g_2(x)-1, \quad x \in \R^d.
\label{eqn:C1}
\end{equation}
Under the assumptions {\bf (A1)} and {\bf (A2)},
(\ref{eqn:var1}) is written as 
\begin{align*}
\var[\bra \Xi, \phi \ket]
&=\widetilde{\rho}\left[
\int_{\R^d} |\phi(x)|^2 dx
+\widetilde{\rho} \int_{\R^d \times \R^d} \phi(x) \overline{\phi(y)}
C(x-y) dx dy \right]
\nonumber\\
&=\widetilde{\rho}\left[
\int_{\R^d} dx \, |\phi(x)|^2
+\widetilde{\rho} 
\int_{\R^d} dz \, 
C(z) \int_{\R^d} dx \, 
\phi(x) \overline{\phi(x-z)} \right], 
\end{align*}
where the integral variables are changed as
$(x, y) \to (x, z)$ with $z=x-y$.
Define
\begin{equation}
\cI_{\phi}(z) := \int_{\R^d} \phi(x) \overline{\phi(x-z)} dx,
\quad \phi \in \cB_{\rm c}(\R^d), \quad z \in \R^d.
\label{eqn:int1}
\end{equation}
This may be called the
{\it intersection integral} of $\phi$ with displacement $z$.
Remark that if $\phi \in \cB_{\rm c}(\R^d)$,
then $\cI_{\phi} \in \cB_{\rm c}(\R^d)$. 
Then Lemma \ref{thm:variance} gives the following.

%%%%%%%%%%%%%%%%%%%%%%%%%%%
\begin{prop}
\label{thm:variance2}
Assume {\bf (A1)} and {\bf (A2)}. 
For $\phi \in \cB_{\rm c}(\R^d)$, the variance 
is given by
\[
\var[\bra \Xi, \phi \ket]
=\widetilde{\rho}\left[ 
\int_{\R^d} |\phi(x)|^2 dx 
+ \widetilde{\rho} \int_{\R^d} \cI_{\phi}(x) C(x) dx \right].
\]
\end{prop}
%%%%%%%%%%%%%%%%%%%%%%%%%%%

For $k=(k^{(1)}, \dots, k^{(d)})$, $x=(x^{(1)}, \dots, x^{(d)}) \in \R^d$,
$k \cdot x := \sum_{\ell=1}^{d} k^{(\ell)} x^{(\ell)}$, and 
with an integrable function $\varphi$ 
the Fourier transform is defined by
\begin{equation}
\widehat{\varphi}(k) =\sF[\varphi](k):=
\int_{\R^d}
e^{\sqrt{-1} k \cdot x} \varphi(x) dx,
\label{eqn:Fourier1}
\end{equation}
and the inverse Fourier transform is given by
\begin{equation}
\varphi(x) =\sF^{-1}[\widehat{\varphi}](x):=
\frac{1}{(2 \pi)^d} \int_{\R^d}
e^{-\sqrt{-1} k \cdot x} \widehat{\varphi}(k) dk.
\label{eqn:Fourier2}
\end{equation}
Note that
$\varphi(-x)=\varphi(x) \iff 
\widehat{\varphi}(-k)=\widehat{\varphi}(k)$.
If $\varphi(x)$ and $\psi(x)$ are square integrable,
then the \textit{Parseval formula} holds,
\begin{equation}
\int_{\R^d} \varphi(x) \overline{\psi(x)} dx
=\frac{1}{(2\pi)^d} \int_{\R^d} \widehat{\varphi}(k) \overline{\widehat{\psi}(k)} dk.
\label{eqn:Parseval}
\end{equation}
By the definition (\ref{eqn:Fourier1}), we have
$\sF[\phi(\cdot -z)](k)
=\widehat{\phi}(k) e^{\sqrt{-1} k \cdot z}$.
Hence, using the Parseval formula (\ref{eqn:Parseval})
for $\phi \in \cB_{\rm c}(\R^d)$, 
(\ref{eqn:int1}) is written as
\[
\cI_{\phi}(z) = \frac{1}{(2 \pi)^d} 
\int_{\R^d} \widehat{\phi}(k) 
\overline{\widehat{\phi}(k) e^{\sqrt{-1} k \cdot z}} dk
= \frac{1}{(2 \pi)^d} 
\int_{\R^d} e^{-\sqrt{-1} k \cdot z} 
|\widehat{\phi}(k)|^2 dk, 
\quad z \in \R^d.
\]
Comparing this with (\ref{eqn:Fourier2}), we see that
\begin{equation}
\widehat{\cI_{\phi}}(k)=|\widehat{\phi}(k)|^2,
\quad \phi \in \cB_{\rm c}(\R^d), \quad k \in \R^d.
\label{eqn:int2}
\end{equation}

We put the third assumption. 
%%%%%%%%%%%%%%%%%%%%%%%%%%%%%%%%%%%%%%%%%
\begin{description}
\item{\bf (A3)} 
Provided $S=\R^d$ with $d \in \N$, 
the total correlation function $C(x), x \in \R^d$ is square integrable, 
and thus so is its Fourier transform $\widehat{C}(k), k \in \R^d$. 
\end{description}
%%%%%%%%%%%%%%%%%%%%%%%%%%%%%%%%%%%%%
Define 
\begin{equation}
\widehat{S}(k)=1+\widetilde{\rho} \widehat{C}(k), \quad
k \in \R^d,
\label{eqn:S}
\end{equation}
which is called the \textit{structure factor}. 
By definition, this is even;
$\widehat{S}(-k)=\widehat{S}(k), k \in \R^d$.
By the Parseval formula (\ref{eqn:Parseval}),
Proposition \ref{thm:variance2} gives the following.
%%%%%%%%%%%%%%%%%%%%%%%%%%%
\begin{prop}
\label{thm:variance_Fourier}
Assume {\bf (A1)}--{\bf (A3)}. 
For $\phi \in \cB_{\rm c}(\R^d)$, the variance 
is given by
\[
\var[\bra \Xi, \phi \ket]
=\frac{\widetilde{\rho}}{(2\pi)^d} 
\int_{\R^d} \widehat{\cI_{\phi}}(k)
\widehat{S}(k) dk.
\]
\end{prop}
%%%%%%%%%%%%%%%%%%%%%%%%%%%

%%%%%%%%%%%%%%%%%%%%%%%%%%%%%%%%%%%%%%%%%%%%%%%%%
\subsection{Bessel functions}
\label{sec:Bessel}
%%%%%%%%%%%%%%%%%%%%%%%%%%%%%%%%%%%%%%%%%%%%%%%%%

The \textit{Bessel function of the
first kind} is defined as \cite{Wat44,NIST10} 
\begin{equation}
J_{\nu}(z)= \left( \frac{z}{2} \right)^{\nu}
\sum_{n=0}^{\infty} (-1)^n \frac{(z/2)^{2n}}{n! \Gamma(\nu+n+1)},
\quad 
z \in \C \setminus (-\infty, 0].
\label{eqn:Bessel1}
\end{equation}
If $\varphi(x)$, 
$x=(x^{(1)}, \dots, x^{(d)}) \in \R^d$ depends only on
the modulus 
$|x| = \sqrt{\sum_{\ell=1}^d (x^{(\ell)})^2}$,
it is said to be \textit{radial}.
The following lemma is well known
(see, for instance, \cite[Section 2.1]{Tor18}).

%%%%%%%%%%%%%%%%%%%%%%%%
\begin{lem}
\label{thm:Bessel1}
If the integrable function $\varphi(x), x \in \R^d$ is 
radial and expressed as
$\varphi(x)=f(r)$ with $r=|x|$, then
its Fourier transform (\ref{eqn:Fourier1}) is
also radial and given by a function of $\kappa:=|k|$ as
\begin{align*}
\widehat{\varphi}(k)
=\widehat{f}(\kappa)
&=(2 \pi)^{d/2} \int_0^{\infty} r^{d-1} 
\frac{J_{(d-2)/2}(\kappa r)}{(\kappa r)^{(d-2)/2}} f(r) dr
\nonumber\\
&= \frac{(2 \pi)^{d/2}}{\kappa^{(d-2)/2}}
\int_0^{\infty} r^{d/2} 
J_{(d-2)/2}(\kappa r) f(r) dr.
\end{align*}
The inverse transform of 
$\widehat{\varphi}(k)=\widehat{f}(\kappa)$ is
given by
\begin{align}
\varphi(x)=f(r)
&=\frac{1}{(2 \pi)^{d/2}} \int_0^{\infty} {\kappa}^{d-1} 
\frac{J_{(d-2)/2}(\kappa r)}{(\kappa r)^{(d-2)/2}} 
\widehat{f}(\kappa) d \kappa
\nonumber\\
&=\frac{1}{(2 \pi)^{d/2} r^{(d-2)/2}}
\int_0^{\infty} {\kappa}^{d/2} 
J_{(d-2)/2}(\kappa r) \widehat{f}(\kappa) d \kappa.
\label{eqn:Bessel_eq2}
\end{align}
\end{lem}
%%%%%%%%%%%%%%%%%%%%%%%%

We will use the following formulas \cite{NIST10,Wat44} 
for an indefinite integral,
\begin{equation}
\int \frac{J_{\nu}(a x)^2}{x^{2\nu-1}} d x
=-\frac{1}{2 (2 \nu-1)}
\frac{J_{\nu-1}(a x)^2+J_{\nu}(a x)^2}{x^{2(\nu-1)}},
\quad \nu \not=1/2, 
\label{eqn:J_formula1}
\end{equation}
and for definite integrals,
\begin{align}
&\int_0^{\infty}
x^{-1} J_{\nu}(ax)^2=\frac{1}{2 \nu},
\label{eqn:J_formula_add1}
\\
&\int_0^{\infty} 
x^{\nu+1} e^{-p^2 x^2} 
J_{\nu}(a x) dx
=\frac{a^{\nu}}{(2p^2)^{\nu+1}} 
e^{-a^2/(4 p^2)}, 
\label{eqn:J_formula2}
\\
&\int_0^{\infty} x e^{-p^2 x^2} J_{\nu}(a x)^2 dx
= \frac{1}{2 p^2} e^{-a^2/(2 p^2)} I_{\nu}(a^2/(2 p^2)),
\label{eqn:J_formula3}
\\
&\int_0^{\infty} x^{-1} e^{-p^2 x^2} J_{\nu}(a x)^2 dx
\nonumber\\
& \qquad 
= \frac{(a/p)^{2\nu}}{2^{2\nu+1} \nu^2 \Gamma(\nu)}
{_{2}F_2}(\nu, \nu+1/2; \nu+1, 2\nu+1; -(a/p)^2), 
\label{eqn:J_formula_add2}
\end{align}
$\Re \nu > -1, \Re p^2 >0$,
where $I_{\nu}$ 
and ${_{2}F_2}$ are defined by
(\ref{eqn:mBessel}) and (\ref{eqn:HG1}), respectively.
Note that (\ref{eqn:J_formula_add2}) is a special case
of the integral formula given in Section 13.32 of \cite{Wat44}
which is expressed using ${_3 F_3}$.
The following asymptotics will be also used \cite{Wat44,NIST10},
\begin{align}
J_{\nu}(x) &\sim 
\sqrt{\frac{2}{\pi x}} 
\left\{
\cos \omega_{\nu}(x) 
\sum_{k=0}^{\infty} (-1)^k \frac{\alpha_{2k}(\nu)}{(2k)! 2^{6k}}x^{-2k}
\right.
\nonumber\\
& \qquad \qquad \qquad 
\left.
-\sin \omega_{\nu}(x) \sum_{k=0}^{\infty} (-1)^k 
\frac{\alpha_{2k+1}(\nu)}{(2k+1)! 2^{3(2k+1)}} x^{-2k-1}
\right\}
\nonumber\\
&\sim 
\sqrt{\frac{2}{\pi x}} 
\cos\omega_{\nu}(x), 
\quad \mbox{as $x \to \infty$},
\label{eqn:J_formula4}
\\
I_{\nu}(x) &\sim
\frac{e^x}{\sqrt{2 \pi x}}
\sum_{k=0}^{\infty} (-1)^k \frac{\alpha_k(\nu)}{k! 2^{3k}} x^{-k},
\quad \mbox{as $x \to \infty$},
\label{eqn:I_formula1}
\end{align}
where
$\omega_{\nu}(x)=x-(2\nu+1)\pi/4$
and $\alpha_k, k \in \N_0:=\{0,1,2, \dots\}$ 
are defined by (\ref{eqn:alpha1}). 

%%%%%%%%%%%%%%%%%%%%%%%%%%%%%%%%%%%%%%%%%%%%%%%%%%%%%%%%%%
%%%  SEC3 %%%%%%%%%%%%%%%%%%%%%%%%%%%%%%%%%%%%%%%%%%%
%%%%%%%%%%%%%%%%%%%%%%%%%%%%%%%%%%%%%%%%%%%%%%%%%%%%%%%%%%
\SSC
{Local Number Variances} \label{sec:number}%%%%%%%%
%%%%%%%%%%%%%%%%%%%%%%%%%%%%%%%%%%%%%%%%%%%%%%%%%
\subsection{General formulas}
\label{sec:general}
%%%%%%%%%%%%%%%%%%%%%%%%%%%%%%%%%%%%%%%%%%%%%%%

An indicator function of a domain 
$\Lambda \subset S$ is defined by
\[
1_{\Lambda}(x) := \begin{cases}
1, \quad & \mbox{if $x \in \Lambda$}, \cr
0, \quad & \mbox{otherwise}.
\end{cases}
\]
Here we consider the case that
$S=\R^d, d \in \N$ and $\Lambda=\B^{(d)}_R$
with $R >0$.
By definition $1_{\B^{(d)}_R}(x)$ is radial and we write
$1_{\B^{(d)}_R}(x)= \chi_{\B^{(d)}_R}(|x|)$.
For $\phi=1_{\B^{(d)}_R}$, the intersection integral 
(\ref{eqn:int1}) becomes
\begin{equation}
\cI_{1_{\B^{(d)}_{R}}}(x)=
\int_{\R^d} 1_{\B^{(d)}_R}(y) 1_{\B^{(d)}_R}(y-x) dy, \quad x \in \R^d. 
\label{eqn:int_v1}
\end{equation}
This is called the 
\textit{intersection volume} of two balls with radius $R$
whose centers are separated by $x$ \cite{TS03,TS06,Tor18}.
By definition $\cI_{1_{\B^{(d)}_{R}}}(x)=0$ for
$x \in \R^d \setminus \B^{(d)}_R$. 

Under the assumptions {\bf (A1)} and {\bf (A2)},
(\ref{eqn:1_correlation}) gives
\begin{equation}
\bE[\Xi(\B^{(d)}_R)]
= \widetilde{\rho} \int_{\R^d} 1_{\B^{(d)}_R}(x) dx
=\vol(\B^{(d)}_R) \widetilde{\rho},
\label{eqn:E1}
\end{equation}
where $\vol(\B^{(d)}_R)$ is given by (\ref{eqn:vol}).

As an application of Lemma \ref{thm:Bessel1},
we have the following.
See, for instance, \cite{KS20,KS19+} for proof.
%%%%%%%%%%%%%%%%%%%%%%%%%%%%%%%%%%
\begin{lem}
\label{thm:ball}
The Fourier transform of $1_{\B^{(d)}_R}(x)$
\[
\widehat{1_{\B^{(d)}_R}}(k)
:= \int_{\R^d} e^{\sqrt{-1} k \cdot x} 1_{\B^{(d)}_R}(x) dx
=\int_{\B^{(d)}_R} e^{\sqrt{-1} k \cdot x} dx
\]
is radial and given as a function of
$\kappa:=|k|$. If we write it as 
$\widehat{1_{\B^{(d)}_R}}(k)
=\widehat{\chi_{\B^{(d)}_R}}(\kappa)$, 
then we have
\[
\widehat{\chi_{\B^{(d)}_R}}(\kappa)
=\frac{(2 \pi)^{d/2}}{\kappa^{(d-2)/2}}
\int_0^R r^{d/2} J_{(d-2)/2}(\kappa r) dr
= (2 \pi)^{d/2}
\left( \frac{R}{\kappa} \right)^{d/2}
J_{d/2}(\kappa R).
\]
\end{lem}
%%%%%%%%%%%%%%%%%%%%%%%%%%

With the relation (\ref{eqn:int2}) the above lemma gives
the Fourier transform of the intersection volume 
(\ref{eqn:int_v1}) as
\begin{equation}
\widehat{\cI_{1_{\B^{(d)}_R}}}(k)
=(2 \pi)^d R^d 
\frac{J_{d/2}(\kappa R)^2}{\kappa^d}
=: \widehat{\cI_{\chi_{\B^{(d)}_R}}}(\kappa), 
\quad k \in \R^d, \quad \kappa=|k|. 
\label{eqn:int_v2}
\end{equation}
The intersection volume (\ref{eqn:int_v1}) is then
obtained as a function of the modulus $r=|x|$ 
by performing the inverse Fourier transform (\ref{eqn:Bessel_eq2}) 
of (\ref{eqn:int_v2}); 
\begin{align*}
\cI_{1_{\B^{(d)}_R}}(x)
&= \sF^{-1} \Big[
\widehat{\cI_{1_{\B^{(d)}_R}}} \Big](x)
\nonumber\\
&= \frac{(2 \pi)^{d/2}}{r^{(d-2)/2}} R^d
\int_0^{\infty} 
\frac{J_{d/2}(\kappa R)^2 J_{(d-2)/2}(\kappa r)}
{\kappa^{d/2}} d \kappa
=: \cI_{\chi_{\B^{(d)}_R}}(r),
\quad r=|x| < 2R.
\end{align*}
By the definition (\ref{eqn:int1}), 
$\cI_{\chi_{\B^{(d)}_R}}(r)=0$ if $r \geq 2R$.
Hence as a corollary of Propositions
\ref{thm:variance2} and \ref{thm:variance_Fourier}
we have the following.

%%%%%%%%%%%%%%%%%%%%%%%%%%%
\begin{cor}
\label{thm:number_fluctuation}
\begin{description}
\item{\rm (i)} 
Assume {\bf (A1)} and {\bf (A2)}. 
Then
\[
\var[\Xi(\B^{(d)}_R)]
=\widetilde{\rho} \left[ \vol(\B^{(d)}_R) 
+ \widetilde{\rho} 
\int_{\R^d} \cI_{\chi_{\B^{(d)}_R}}(|x|) C(x) dx
\right], 
\]
where $C(x)$ is the total correlation function (\ref{eqn:C1}).
When $C(x)$ is radial and written as $C(x)=c(r)$ 
with $r=|x|$, then
\begin{equation}
\var[\Xi(\B^{(d)}_R)]
=\widetilde{\rho} \left[ \vol(\B^{(d)}_R) 
+ \frac{2 \pi^{d/2} \widetilde{\rho}}{\Gamma(d/2)}
\int_0^{2R} \cI_{\chi_{\B^{(d)}_R}}(r) c(r) r^{d-1} dr
\right].
\label{eqn:NF_real2}
\end{equation}
\item{\rm (ii)} 
Assume {\bf (A1)}--{\bf (A3)}. Then
\begin{equation}
\var[\Xi(\B^{(d)}_R)]
=\frac{\widetilde{\rho}}{(2 \pi)^d} 
\int_{\R^d} \widehat{\cI_{1_{\B^{(d)}_R}}}(|k|) \widehat{S}(k) dk
= \widetilde{\rho} R^d
\int_{\R^d} \frac{J_{d/2}(|k| R)^2}{|k|^d} 
\widehat{S}(k) dk,
\label{eqn:number_fluctuation1}
\end{equation}
where $\widehat{S}(k)$ is the structure factor (\ref{eqn:S}).
When $\widehat{S}(k)$ is radial and written as
$\widehat{S}(k)=\widehat{s}(\kappa)$ with
$\kappa=|k|$, 
then
\begin{equation}
\var[\Xi(\B^{(d)}_R)]
=\frac{2 \pi^{d/2} \widetilde{\rho}}{\Gamma(d/2)} R^d
\int_0^{\infty} \frac{J_{d/2}(\kappa R)^2}{\kappa}
\widehat{s}(\kappa) d \kappa.
\label{eqn:number_fluctuation2}
\end{equation}
\end{description}
\end{cor}
%%%%%%%%%%%%%%%%%%%%%%%%%%%
\noindent \textit{Proof} \,
The formulas (\ref{eqn:NF_real2}) and
(\ref{eqn:number_fluctuation2}) are obtained, if we use the 
polar coordinate expression of the 
Lebesgue measure for radial functions; 
$dx=r^{d-1} \sigma_{d-1} d r$
with $\sigma_{d-1}=2 \pi^{d/2}/\Gamma(d/2)$.
\qed
%%%%%%%%%%%%%%%%%%%%%%%%

%%%%%%%%%%%%%%%%%%%%%%%%%%%%%%%%%%%%%%%%%%%%%%%%%
\subsection{Determinantal point processes}
\label{sec:DPP}
%%%%%%%%%%%%%%%%%%%%%%%%%%%%%%%%%%%%%%%%%%%%%%%

Determinantal point process (DPP) 
is defined as follows 
\cite{Sos00,ST03a,ST03b,HKPV06,HKPV09,KS19+}. 
%%%%%%%%%%%%%%%%%%%%%%%%%%%
\begin{df}
A simple point process $\Xi$ 
on $(S, \cB_{\rm c}(S), \lambda)$ is said to be a DPP with 
a measurable kernel $K : S \times S \to \C$, if it 
satisfies the assumption {\bf (A1)} so that 
the correlation functions 
with respect to $\lambda$ are given by
\[
\rho_n(x_1, \dots, x_n) = \det_{1 \leq i, j \leq n}
[ K(x_i, x_j) ]
\quad \mbox{for every $n \in \N$
and any $x_1, \dots, x_n \in S$}.
\]
The integral kernel $K$ is called the
correlation kernel. 
The DPP is specified by the triplet $(\Xi, K, \lambda)$.
\end{df}
%%%%%%%%%%%%%%%%%%%%%%%%%%

If the point process $\Xi$ is a DPP, then
(\ref{eqn:1_correlation}) and 
(\ref{eqn:var1}) in Lemma \ref{thm:variance} are
given by
\begin{align*}
\bE[\bra \Xi, \phi \ket]
&= \int_{S} \phi(x) K(x, x) \lambda(dx),
\nonumber\\
\var[\bra \Xi, \phi \ket]
&= \frac{1}{2} \int_{S \times S} 
|\phi(x)-\phi(y)|^2 K(x, y) K(y,x) \lambda^{\otimes 2}(dx dy),
\quad \phi \in \cB_{\rm c}(S).
\end{align*}
In particular, when $\phi=1_{\Lambda}$ for
a bounded domain $\Lambda \subset S$, 
the above give the following,
\begin{align}
\bE[\bra \Xi(\Lambda) \ket]
&= \int_{\Lambda} K(x, x) \lambda(dx),
\nonumber\\
\var[\bra \Xi(\Lambda) \ket]
&= \int_{\Lambda} 
\int_{S \setminus \Lambda}
K(x, y) K(y,x) \lambda(dx) \lambda(dy).
\label{eqn:DPP_special}
\end{align}

Provided that $S=\R^d$, $d \in \N$, or
$S=\C^D \simeq \R^d$ with $d=2D$, $D \in \N$, 
we put additional assumptions.

%%%%%%%%%%%%%%%%%%%%%%%%%%%%%%%%%%%%%%%%%
\begin{description}
\item{\bf (DPP)} 
The point process $\Xi$ is a DPP on $S$, $(\Xi, K, \lambda)$, 
and the following are satisfied. 
\begin{description}
\item{(i)} \,
The correlation kernel is hermitian,
\[
\overline{K(x,y)}=K(y, x), \quad x, y \in S.
\]
\item{(ii)} \,
The reference measure is given in the form
$\lambda(dx)=\ell(x) dx, x \in S$, and
\[
K(x, x) \ell(x) = \mbox{constant} 
=: \widetilde{\rho} \quad \forall x \in S.
\]
\item{(iii)} \,
There is a measurable even function $C(x)=C(-x)$, 
$x \in S$ 
such that 
\[
C(x-y)=-\frac{|K(x,y)|^2}{K(x,x) K(y,y)},
\quad x, y \in S.
\]
\end{description}
\end{description}
%%%%%%%%%%%%%%%%%%%%%%%%%%%%%%%%%%%%%

%%%%%%%%%%%%%%%%%%%%%%%%%%%
\begin{cor}
\label{thm:number_fluctuation_DPP}
Assume {\bf (DPP)} and {\bf (A3)}.
Then the assertions of Corollary \ref{thm:number_fluctuation} (ii)
hold.
\end{cor}
%%%%%%%%%%%%%%%%%%%%%%%%%%%

%%%%%%%%%%%%%%%%%%%%%%%%%%%%%%
\subsection{Heisenberg family of DPPs on $\C^D$}
\label{sec:Heisenberg}
%%%%%%%%%%%%%%%%%%%%%%%%%%%%%%%%%

%%%%%%%%%%%%%%%%%%%%%%%%%%%%%%
\begin{lem}
\label{thm:H_DPP}
The Heisenberg family of DPPs,  
$(\Xi_{\sH_D}, K_{\sH_D}, \lambda_{\rN(0, 1; \C^D)})$
on $\C^D$, 
$D \in \N$ satisfies {\bf (DPP)} for
$S = \C^D \simeq \R^{d}$ with $d=2D$,
\begin{equation}
\widetilde{\rho} =\frac{1}{\pi^D},
\label{eqn:rhotilde_H1}
\end{equation}
and
\begin{equation}
C(x)=c(|x|) = -e^{-|x|^2},
\label{eqn:c_H1}
\end{equation}
where $|x|^2 =|x_{\rR}|^2+|x_{\rI}|^2, x \in \C^D$. 
\end{lem}
%%%%%%%%%%%%%%%%%%%%%%
\noindent \textit{Proof} \,
By Definition \ref{thm:HeisenbergDPP},
$\rho_1(x)=K_{\sH_D}(x,x)=e^{|x|^2}$,
$x \in \C^D$. Then
$\widetilde{\rho}= e^{|x|^2} e^{-|x|^2}/\pi^D
=1/\pi^D$ proving (\ref{eqn:rhotilde_H1}). 
Since $K_{\sH_D}(x,y)=e^{x \cdot \overline{y}}$, we see that
\begin{align*}
C(x-y) &=-\frac{|K_{\sH_D}(x,y)|^2}{K_{\sH_D}(x,x) K_{\sH_D}(y,y)}
=-e^{x \cdot \overline{y}+\overline{x} \cdot y-|x|^2-|y|^2}
\nonumber\\
&=-e^{-|x-y|^2}=: c(|x-y|), 
\end{align*}
which proves (\ref{eqn:c_H1}).
\qed
%%%%%%%%%%%%%%%%%%%%%%%%

Since $\C^D \simeq \R^d$ with $d=2D$, 
a disk centered at the origin with radius $R$ on
$\C^D$, $\{ x \in \C^D : |x| < R\}$,
is identified with $\B^{(2D)}_R$ in $\R^{2D}$. 

%%%%%%%%%%%%%%%%%%%%%
\begin{prop}
\label{thm:H_DPP2}
For the Heisenberg family of DPPs,  
$(\Xi_{\sH_D}, K_{\sH_D}, \lambda_{\rN(0, 1; \C^D)})$ 
on $\C^D$,
$D \in \N$, the following hold,
\begin{align}
\bE[\Xi_{\sH_D}(\B^{(2D)}_R)]
&= \frac{R^{2D}}{D!},
\label{eqn:E_H1}
\\
\var[\Xi_{\sH_D}(\B^{(2D)}_R)]
&=\frac{2 R^{2D}}{(D-1)!} 
\int_0^{\infty} 
\frac{J_D(\kappa R)^2}{\kappa}
(1-e^{-\kappa^2/4}) d \kappa,
\quad R>0.
\label{eqn:Var_H1}
\end{align}
\end{prop}
%%%%%%%%%%%%%%%%%%%%%%
\noindent \textit{Proof} \,
Combining (\ref{eqn:E1}) with 
(\ref{eqn:vol}), and (\ref{eqn:rhotilde_H1}), 
(\ref{eqn:E_H1}) is proved. 
Since $C(x)$ is radial as given by (\ref{eqn:c_H1})
in Lemma \ref{thm:H_DPP}, 
Lemma \ref{thm:Bessel1} determines its Fourier transform
$\widehat{C}(k)$ as a radial function of 
$\kappa=|k|$ as
\[
\widehat{c}(\kappa)
=\frac{(2 \pi)^D}{\kappa^{D-1}}
\int_0^{\infty} r^D J_{D-1}(\kappa r) c(r) dr
=- \frac{(2 \pi)^D}{\kappa^{D-1}}
\int_0^{\infty} r^D J_{D-1}(\kappa r) e^{-r^2} dr.
\]
We use the integral formula (\ref{eqn:J_formula2}) with
$\nu=D-1, a=\kappa, p=1$ and obtain
$\widehat{c}(\kappa)=-\pi^D e^{-\kappa^2/4}$.
In this sense, the assumption {\bf (A3)} is satisfied 
for the present systems on $S=\C^D, D \in \N$. 
With (\ref{eqn:rhotilde_H1}) of Lemma \ref{thm:H_DPP}, 
(\ref{eqn:S}) gives
\[
\widehat{s}(\kappa)=1+\frac{1}{\pi^D} (- \pi^D e^{-\kappa^2/4})
=1-e^{-\kappa^2/4}.
\]
Then Corollary \ref{thm:number_fluctuation_DPP} proves
(\ref{eqn:Var_H1}). 
The proof is hence complete.
\qed

%%%%%%%%%%%%%%%%%%%%%%%%%%%%%%%%%%%%%%%%%%%%%%%%%%%%%%%%%%
%%%  SEC4 %%%%%%%%%%%%%%%%%%%%%%%%%%%%%%%%%%%%%%%%%%%
%%%%%%%%%%%%%%%%%%%%%%%%%%%%%%%%%%%%%%%%%%%%%%%%%%%%%%%%%%
\SSC
{Proofs of Main Results} \label{sec:proofs}%%%%%%%%
%%%%%%%%%%%%%%%%%%%%%%%%%%%%%%%%%%%%%%%%%%%%

%%%%%%%%%%%%%%%%%%%%%%%%%%%%%%%%%%%%%%%%%%%%%%%%%%%%%%%%%%
\subsection
{Proof of Proposition \ref{thm:main1}} \label{sec:proof1}%%%%%%%%
%%%%%%%%%%%%%%%%%%%%%%%%%%%%%%%%%%%%%%%%%%%%

For $n \in \N$, 
consider the integral
\begin{align}
A_n(R) &:= \int_0^{\infty} 
\frac{J_n(\kappa R)^2}{\kappa}
(1-e^{-\kappa^2/4}) d \kappa
\nonumber\\
&=\int_0^{\infty} 
\frac{J_n(\kappa R)^2}{\kappa^{2n-1}}
\kappa^{2(n-1)} (1-e^{-\kappa^2/4}) d \kappa.
\label{eqn:AnR}
\end{align}
By (\ref{eqn:J_formula1}),  we can perform the 
partial integration as
\begin{align}
A_n(R) &= \left[
-\frac{1}{2(2n-1)} 
[J_{n-1}(\kappa R)^2 + J_{n}(\kappa R)^2 ]
(1-e^{-\kappa^2/4}) \right]_0^{\infty}
\nonumber\\
&+ \frac{1}{2(2n-1)} 
\int_0^{\infty} \frac{J_{n-1}(\kappa R)^2 + J_{n}(\kappa R)^2}
{\kappa^{2(n-1)}}
\left[ \frac{d}{d \kappa}
\{\kappa^{2(n-1)} (1-e^{-\kappa^2/4}) \} 
\right] d \kappa.
\label{eqn:Id1}
\end{align}
The asymptotic formula (\ref{eqn:J_formula4}) implies
$J_n(\kappa R)^2 + J_{n-1}(\kappa R)^2 
\sim 2/(\pi \kappa R) \to 0$
as $\kappa R \to \infty$,
and hence the first term
in the RHS of (\ref{eqn:Id1}) vanishes.
Since
\[
\frac{d}{d \kappa}
\{\kappa^{2(n-1)} (1-e^{-\kappa^2/4})\}
=2(n-1) \kappa^{2(n-1)-1}(1-e^{-\kappa^2/4})
+\frac{1}{2} \kappa^{2(n-1)+1} e^{-\kappa^2/4},
\]
(\ref{eqn:Id1}) is written as
\begin{align*}
A_n(R) &=\frac{n-1}{2n-1} A_n(R)+\frac{n-1}{2n-1} A_{n-1}(R)
\nonumber\\
& \quad
+ \frac{1}{4(2n-1)} \int_0^{\infty} \kappa e^{-\kappa^2/4}
[J_{n-1}(\kappa R)^2+J_{n}(\kappa R)^2 ] d \kappa
\nonumber\\
&=\frac{n-1}{2n-1} A_n(R)+\frac{n-1}{2n-1} A_{n-1}(R)
+\frac{1}{2(2n-1)} e^{-2 R^2} [I_{n-1}(2R^2) + I_{n}(2R^2)],
\end{align*}
where (\ref{eqn:J_formula3}) was used.
This gives the following recurrence relation
\[
n A_n(R)-(n-1) A_{n-1} 
=\frac{e^{-2R^2}}{2} [I_{n-1}(2R^2)+I_{n}(2R^2)],
\quad n \in \N.
\]
By taking the summation with respect to $n$ 
from 1 to $D \in \N$, we have
\begin{align*}
D A_D(R) &=\frac{e^{-2R^2}}{2} 
\sum_{n=1}^D [I_{n-1}(2R^2) + I_{n}(2R^2)]
\nonumber\\
&= \frac{e^{-2R^2}}{2}
\left[ I_0(2R^2) + 2 \sum_{n=1}^{D-1} I_n(2R^2) +I_D(2R^2)
\right].
\end{align*}
With (\ref{eqn:Var_H1}) this proves 
the assertion (\ref{eqn:main1}). 

%%%%%%%%%%%%%%%%%%%%%%%%%%%%%%%%%%%%%%%%%%%%%%%%%%%%%%%%%%
\subsection
{Proof of (\ref{eqn:HG2}) in Remark 2} \label{sec:HG}
%%%%%%%%%%%%%%%%%%%%%%%%%%%%%%%%%%%%%%%%%%%%%%%%%%%%

Consider (\ref{eqn:AnR}) with $n=D \in \N$ and set
\[
A_D(R)
= A_D^{(1)}(R)-A_D^{(2)}(R)
\]
with
\[
A_D^{(1)}(R)=\int_0^{\infty} \frac{J_D(\kappa R)^2}{\kappa} d \kappa,
\quad
A_D^{(2)}(R)=\int_0^{\infty} \frac{J_D(\kappa R)^2}{\kappa} 
e^{-\kappa^2/4} d \kappa.
\]
The integral formulas (\ref{eqn:J_formula_add1}) and (\ref{eqn:J_formula_add2})
give
$A_D^{(1)}(R)=1/(2D)$ and
\[
A_D^{(2)}(R)
=\frac{(2R)^{2D}}{2^{2D+1} D^2 \Gamma(D)}
{_{2}F_2}(D, D+1/2; D+1, 2D+1; -(2R)^2).
\]
Putting the above results into (\ref{eqn:Var_H1}), 
the formula (\ref{eqn:HG2}) is obtained.

%%%%%%%%%%%%%%%%%%%%%%%%%%%%%%%%%%%%%%%%%%%%%%%%%%%%%%%%%%
\subsection
{Proof of Theorem \ref{thm:main2}} \label{sec:proof2}
%%%%%%%%%%%%%%%%%%%%%%%%%%%%%%%%%%%%%%%%%%%%%%%%%%%%

Apply the asymptotic formula of the modified Bessel functions
(\ref{eqn:I_formula1}) with (\ref{eqn:alpha1}).
Then (\ref{eqn:main1}) in Proposition \ref{thm:main1}
combined with (\ref{eqn:E_H1}) in Proposition \ref{thm:H_DPP2}
gives
\[
\frac{\var[\Xi_{\sH_D}(\B^{(2D)}_R)]}
{\bE[\Xi_{\sH_D}(\B^{(2D)}_R)]}
\sim \frac{1}{2 \sqrt{\pi}} R^{-1}
\sum_{k=0}^{\infty} (-1)^k \frac{\beta_k(D)}{k! 2^{4k}} R^{-2k},
\]
where
\[
\beta_k(D)
:=\alpha_k(0)+2 \sum_{n=1}^{D-1} \alpha_k(n) + \alpha_k(D),
\quad k \in \N_0.
\]
Since $\alpha_0(n) \equiv 1$ by the definition (\ref{eqn:alpha1}),
$\beta_0(D)=1+2(D-1)+1=2D$,
and (\ref{eqn:C0}) is proved.
For $k \in \N$, (\ref{eqn:alpha1}) gives
\begin{align*}
\alpha_k(n+1) &= \prod_{\ell=-k+1}^k \{2(n+1)+2 \ell-1\}
= \prod_{\ell'=-k+2}^{k+1} (2n+2 \ell'-1)
\nonumber\\
&= \frac{2n+2k+1}{2n-2k+1} \alpha_k(n),
\quad n \in \N_0.
\end{align*}
This equality is rewritten as
\[
\alpha_k(n)+\alpha_k(n+1)
=\frac{2}{2k+1} [
(n+1) \alpha_k(n+1)- n \alpha_k(n) ],
\quad n \in \N_0.
\]
If we take summation of the above from $n=0$ to $n=D-1$, then we have
\[
\sum_{n=0}^{D-1} \{ \alpha_k(n)+\alpha_k(n+1)\}
=\frac{2D}{2k+1} \alpha_k(D).
\]
This implies
$\beta_k(D)=\{2D \alpha_k(D)\}/(2k+1)$
and (\ref{eqn:asym}) is concluded. 
Then the proof is complete.

%%%%%%%%%%%%%%%%%%%%%%%%%%%%%%%%%%%%%%%%%%%%%%%%%%%%%%%%%%
\subsection
{Proof of (\ref{eqn:polyD1}) in Remark 5} \label{sec:polyD}
%%%%%%%%%%%%%%%%%%%%%%%%%%%%%%%%%%%%%%%%%%%%%%%%%%%%

As shown in Appendix \ref{sec:HeisenbergGroup},
$\{\varphi_n\}_{n \in \N_0^D}$ with (\ref{eqn:phi1})
gives a complete orthonormal system for the Bargmann--Fock space
$\cF_D$ defined by (\ref{eqn:BF_space}).
For 
$n=(n^{(1)}, \dots, n^{(D)})$, 
$m=(m^{(1)}, \dots, m^{(D)}) \in \N_0^D$ and $R >0$, let
\begin{equation}
K_{\Delta^{(D)}_R}(n,m)
:= \int_{\Delta^{(D)}_R} \overline{\varphi_n(x)} \varphi_m(x)
\lambda_{\rN(0,1; \C^D)}(dx),
\label{eqn:K_dual}
\end{equation}
and consider the DPP
$\Xi_{\Delta^{(D)}_R}$ on $\N_0^D$ whose
correlation kernel is given by (\ref{eqn:K_dual}). 
By the general theory of the \textit{duality relations} between DPPs
(Theorem 2.6 in \cite{KS19+}), the following equality holds,
\begin{equation}
\bP(\Xi_{\sH_D}(\Delta^{(D)}_R)=k)
=\bP(\Xi_{\Delta^{(D)}_R}(\N_0^D)=k)
\quad \forall k \in \N_0.
\label{eqn:dual}
\end{equation}

By (\ref{eqn:phi1}), we can show that \cite{KS19+,Shi15}
\[
K_{\Delta^{(D)}_R}(n, m)
=\delta_{n m} \prod_{\ell=1}^D p_{n^{(\ell)}}(R), \quad
n, m \in \N_0^D,
\]
where
\[
p_{k}(R):=\int_0^{R^2} \frac{u^k e^{-u}}{k !} du
=\sum_{j=k+1}^{\infty}
\frac{R^{2j} e^{-R^2}}{j!}, \quad
k \in \N_0.
\]
Then the DPP $(\Xi_{\Delta^{(D)}_R}, K_{\Delta^{(D)}_R})$
on $\N_0^D$ is the product measure
$\bigotimes_{\ell=1}^D \bigotimes_{n^{(\ell)} \in \N_0} 
\mu^{\rm Bernoulli}_{p_{n^{(\ell)}}(R)}$
under the natural identification between $\{0, 1\}^{\N_0^D}$
and the multivariate power set of $\N_0^D$, where
$\mu^{\rm Bernoulli}_{p}$ denotes the
Bernoulli measure of probability $p \in [0, 1]$. 

If we introduce a series of random variables
$Y^{(R)}_{n^{(\ell)}} \in \{0, 1\}$, $n^{(\ell)} \in \N_0$,
$\ell =1, \dots, D$, which are mutually independent
and $Y^{(R)}_{n^{(\ell)}} \sim \mu^{\rm Bernoulli}_{p_{n^{(\ell)}}(R)}$.
then the duality relation (\ref{eqn:dual}) implies the
equivalences in distribution,
\[
\Xi_{\sH_D}(\Delta^{(D)}_R)
\dis= \Xi_{\Delta^{(D)}_R}(\N_0^D)
\dis= \sum_{n \in \N_0^D} 
\prod_{\ell=1}^D Y^{(R)}_{n^{(\ell)}}.
\]
Then we have
\begin{align*}
&\bE[\Xi_{\sH_D}(\Delta^{(D)}_R)]
=\sum_{n \in \N_0^D} \prod_{\ell=1}^D p_{n^{(\ell)}}(R)
=\left( \sum_{k=0}^{\infty} p_k(R) \right)^D,
\nonumber\\
&\var[\Xi_{\sH_D}(\Delta^{(D)}_R)]
= \var \left[
\sum_{n \in \N_0^D} \prod_{\ell=1}^D Y^{(R)}_{n^{(\ell)}} \right]
=  \sum_{n \in \N_0^D} \var \left[
\prod_{\ell=1}^D Y^{(R)}_{n^{(\ell)}} \right]
\nonumber\\
&\quad = \sum_{n \in \N_0^D}
\left[ \prod_{\ell=1}^{D} p_{n^{(\ell)}} (R)
-\left( \prod_{\ell=1}^{D} 
p_{n^{(\ell)}}(R) \right)^2 \right]
=\left( \sum_{k=0}^{\infty} p_k(R) \right)^D
-\left( \sum_{k=0}^{\infty} p_k(R)^2 \right)^D,
\end{align*}
and hence
\[
\frac{\var[\Xi_{\sH_D}(\Delta^{(D)}_R)]}
{\bE[\Xi_{\sH_D}(\Delta^{(D)}_R)]}
=1 - 
\left(\frac{\sum_{k=0}^{\infty} p_k(R)^2}
{\sum_{k=0}^{\infty} p_k(R)} \right)^D. 
\]
When $D=1$, the above gives
\begin{align*}
\frac{\var[\Xi_{\sH_1}(\Delta^{(1)}_R)]}
{\bE[\Xi_{\sH_1}(\Delta^{(1)}_R)]}
&= 1- \frac{\sum_{k=0}^{\infty} p_k(R)^2}
{\sum_{k=0}^{\infty} p_k(R)}
\nonumber\\
&= \frac{\var[\Xi_{\sH_1}(\B^{(2)}_R)]}
{\bE[\Xi_{\sH_1}(\B^{(2)}_R)]},
\end{align*}
where we have used the fact that 
$\Delta_R^{(1)}=\D_R \subset \C$ is identified
with $\B^{(2)}_R \subset \R^2$.
Hence the first equality of (\ref{eqn:polyD1}) is proved.
The second equality of (\ref{eqn:polyD1}) 
is derived by the asymptotic expansion
(\ref{eqn:asym}) for $D=1$ in Theorem \ref{thm:main2}.

\vskip 1cm
%%%%%%%%%%%%%%%%%%%%%%%%%%%%%%%%%%%%%%%%(
\noindent{\bf Acknowledgements} \,
%%%%%%%%%%%%%%%%%%%%%%%%%%%%%%%%%%%%%%%%%%%%%%%%%%%%%
%%%%%%%%%%%%%%%%%%%%%%%%%%%%%%%%%%%%%%%%%%%%%%%%%%
The present authors would like to thank Shinji Koshida
for useful comments on the manuscript.
This work was supported by
the Grant-in-Aid for Scientific Research (C) (No.19K03674),
(B) (No.18H01124), and
(S) (No.16H06338)
of Japan Society for the Promotion of Science.
It was also supported by 
the Research Institute for Mathematical Sciences, 
an International Joint Usage/Research Center located 
in Kyoto University.
%%%%%%%%%%%%%%%%%%%%%%%%%%%%%%%%%%%%%%%%%%

%%%%%%%%%%%%%%%%%%%%%%%%%%%%%%%%%%%%%%%%%%%%%%%%%%%%%%%%%%%%
%%%%%%%%%% APPENDICES %%%%%%%%%%%%%%%%%%%%%%%%%%%%%%%%%%%%%%
%%%%%%%%%%%%%%%%%%%%%%%%%%%%%%%%%%%%%%%%%%%%%%%%%%%%%%%%%%%%
\appendix
%%%%%%%%%%%%%%%%%%%%%%%%%%%%%%%%%%%%%%%%%%%%%%%%%%%%%%%%%%
%%%  Appendix A %%%%%%%%%%%%%%%%%%%%%%%%%%%%%%%%%%%%%%%%%%%
%%%%%%%%%%%%%%%%%%%%%%%%%%%%%%%%%%%%%%%%%%%%%%%%%%%%%%%%%%
\SSC{Representations of the Heisenberg Group}
\label{sec:HeisenbergGroup}
%%%%%%%%%%%%%%%%%%%%%%%%%%%%%%%%%%%%%%%%%%%%%%%%%%%%%%%%%%

Following \cite{Fol89,Ste93,Gro01}, we briefly review
the representation theory of the \textit{Heisenberg group}
in order to explain the reason why we call the
DPPs defined by Definition \ref{thm:HeisenbergDPP}
the \textit{Heisenberg family} of DPPs.

Consider the classical and quantum kinetics of a single particle
moving in $\R^D, D \in \N$.
We note that, if $D=3k, k \in \N$, this 
represents a $k$-particle system in 
the three dimensional Euclidean space.
The \textit{phase space} is given by $\R^{2D}$ with
coordinates 
$(p, q)=(p_1, \dots, p_D, q_1, \dots, q_D)$.
In order to describe the 
\textit{Heisenberg Lie algebra} $\sh_D$, we consider
$\R^{2D+1}$ with coordinates
$(p, q, \tau)=(p_1, \dots, p_D, q_1, \dots, q_D, \tau)$, 
in which a Lie bracket is given by
\[
[(p,q,\tau), (p', q', \tau')]
=(0, 0, p \cdot q'-q  \cdot p')
=(0, 0, [(p,q), (p', q')]).
\]
The symplectic form of the Lie bracket 
$[(p,q), (p',q')]=p \cdot q'-q \cdot p'$ comes from
the Poisson bracket in the classical mechanics
and the commutator $[A, B] := AB-BA$ in quantum mechanics.
The \textit{Heisenberg group} $\sH_D$ is the Lie group
on $\R^{2D+1}$ satisfying the group law
$Z Z'=Z+Z'+\frac{1}{2}[Z, Z']$, $Z, Z' \in \R^{2D+1}$; that is,
\[
(p, q, \tau) (p', q', \tau')
=\left(p+p', q+q', \tau+\tau'+ \frac{1}{2}(p \cdot q'-q \cdot p') \right).
\]

Let $L^2(\R^D)$ be the set of square integrable functions on
$\R^D$, where the inner product is given by
\[
\bra f, g \ket_{L^2(\R^D)}
:= \int_{\R^D} f(\zeta) \overline{g(\zeta)} d \zeta,
\quad f, g \in L^2(\R^D)
\]
with the norm
$\|f \|_{L^2(\R^D)}:=\sqrt{\bra f, f \ket_{L^2(\R^D)}}$,
$f \in L^2(\R^D)$, 
where 
$\zeta=(\zeta^{(1)}, \dots, \zeta^{(D)}) \in \R^D$ and
$d \zeta$ denotes the Lebesgue measure on $\R^D$.
For a smooth function $f$,
we introduce operators $X^{(\ell)}$ and $\cD^{(\ell)}$ defined by
\[
(X^{(\ell)} f)(\zeta)=\zeta^{(\ell)} f(\zeta), \quad
(\cD^{(\ell)} f)(\zeta)
=\frac{1}{2 \sqrt{-1}} \frac{\partial f}{\partial \zeta^{(\ell)}} (\zeta),
\quad \ell=1, \dots, D.
\]
They satisfy the commutation relations
\[
[X^{(\ell)}, \cD^{(\ell')}]=\frac{\sqrt{-1}}{2} \delta_{\ell \ell'}, \quad
\ell, \ell' = 1, \dots, D.
\]
Note that the above will represent the
\textit{canonical commutation relations} in quantum mechanics,
$[Q^{(\ell)}, P^{(\ell')}]=\sqrt{-1} \hbar \delta_{\ell \ell'}$.
Here we should claim that the value of the Planck constant
$\hbar$ is specially chosen to be 1/2.
(This choice enables us to have the
equality (\ref{eqn:M4}) below with
the complex standard normal distribution
$\lambda_{\rN(0,1; \C^D)}$ on $\C^D$
defined by (\ref{eqn:lambdaN}).)
We consider a map from $\sH_D$ to the group of
unitary operators acting on $L^2(\R^D)$ defined by
\[
\rho(p, q, \tau)=e^{2 \sqrt{-1} (p \cdot \cD + q X+ \tau I)},
\]
where $\cD:=(\cD^{(1)}, \dots, \cD^{(D)})$,
$X:=(X^{(1)}, \dots, X^{(D)})$ and $I$ denotes the identity operator.
We can show that
\begin{equation}
\rho(p, q, \tau) f(\zeta)
=e^{2 \sqrt{-1} (\tau + q \cdot \zeta + p \cdot q/2)} 
f(\zeta+p),
\quad f \in L^2(\R^D).
\label{eqn:Sch_rep1}
\end{equation}
The map $\rho$ is called the \textit{Schr\"{o}dinger representation}
of $\sH_D$. 
The kernel of $\rho$ is $\{(0, 0, k \pi) : k \in \Z \} $,
since $e^{2 \pi k \sqrt{-1}}=1, k \in \Z$.
The \textit{reduced Heisenberg group}
$\sH_D^{\rm red}$ is defined by
$\sH_D^{\rm red}:=\sH_D /\{(0, 0, k \pi): k \in \Z\}$.

We calculate the matrix coefficients of $\rho(p, q, \tau)$ 
at $(f, g) \in (L^2(\R^D))^2$ and obtain the expression, 
\begin{align}
M_{f, g}(p,q,\tau)
&:= \bra \rho(p, q, \tau) f, g \ket_{L^2(\R^D)}
\nonumber\\
&=e^{2 \sqrt{-1} \tau} 
\int_{\R^D} e^{2 \sqrt{-1} q \cdot \zeta}
f \left( \zeta + \frac{p}{2} \right)
\overline{g \left(\zeta-\frac{p}{2} \right)} d \zeta,
\quad f, g \in L^2(\R^2), 
\label{eqn:M1}
\end{align}
which is called the \textit{Fourier--Wigner transform} \cite{Fol89}.
For $f_1, f_2, g_1, g_2 \in L^2(\R^D)$, the inner product
of $M_{f_1, g_1}$ and $M_{f_2, g_2}$ in 
$L^2(\R^{2D})$ is calculated and the following equality is obtained,
\begin{align}
\bra M_{f_1, g_1}, M_{f_2, g_2} \ket_{L^2(\R^{2D})}
&:= \int_{\R^D} d p \int_{\R^D} dq \,
M_{f_1, g_1}(p, q, \tau) 
\overline{M_{f_2, g_2}(p, q, \tau)}
\nonumber\\
&= \pi^D \bra f_1, f_2 \ket_{L^2(\R^D)}
\overline{ \bra g_1, g_2 \ket_{L^2(\R^D)}}.
\label{eqn:M2}
\end{align}
If we put
$g_1=g_2=G$ with (\ref{eqn:G1})
and define the complex variables
\begin{equation}
x=(x^{(1)}, \dots, x^{(D)})
:= p + \sqrt{-1} q
=(p^{(1)}+\sqrt{-1} q^{(1)}, \dots, p^{(D)}+\sqrt{-1} q^{(D)}) 
\in \C^D,
\label{eqn:x_pq}
\end{equation}
then (\ref{eqn:M1}) and (\ref{eqn:M2}) become
\[
M_{f, G}(p, q, \tau)
=e^{2 \sqrt{-1} \tau} \sB[f](x)
\frac{e^{-|x|^2/2}}{\pi^{D/2}},
\]
and
\begin{align}
\bra M_{f_1, G}, 
M_{f_2, G} \ket_{L^2(\R^{2D})}
&= \bra f_1, f_2 \ket_{L^2(\R^D)}
\nonumber\\
&= \bra \sB[f_1], \sB[f_2]
\ket_{L^2(\C^D, \lambda_{\rN(0, 1; \C^D)})},
\quad f_1, f_2 \in L^2(\R^D),
\label{eqn:M4}
\end{align}
with
\begin{equation}
\sB[f](x)
:= \left( \frac{2}{\pi} \right)^{D/4}
\int_{\R^D} f(\zeta) e^{2 \zeta \cdot x-\zeta^2-x^2/2} d \zeta,
\quad f \in L^2(\R^D),
\label{eqn:BT}
\end{equation}
which is called the \textit{Bargmann transform}.
Here the measure $\lambda_{\rN(0,1; \C^D)}$ on $\C^D$
is defined by (\ref{eqn:lambdaN}), 
\[
\bra F_1, F_2 \ket_{L^2(\C^D, \lambda_{\rN(0, 1; \C^D)})}
:= \int_{\C^D} F_1(x) \overline{F_2(x)} 
\lambda_{\rN(0,1; \C^D)}(dx),
\]
and 
$\| F \|_{L^2(\C^D, \lambda_{\rN(0, 1; \C^D)})}
:=\sqrt{\bra F, F \ket_{L^2(\C^D, \lambda_{\rN(0, 1; \C^D)})}}$.
For $f \in L^2(\R^D)$, the integral of (\ref{eqn:BT}) 
converges uniformly for $x$ in any compact subset of $\C^D$,
and hence $\sB[f]$ is an entire function on $\C^D$.
The \textit{Bargmann--Fock space} $\cF_D$ is
defined by
\begin{equation}
\cF_D := \left\{F : \mbox{$F$ is entire on $\C^D$ and
$\| F \|_{L^2(\C^D, \lambda_{\rN(0, 1; \C^D)})} < \infty$} \right\}.
\label{eqn:BF_space}
\end{equation}
Then (\ref{eqn:M4}) implies that
the Bargmann transform is an isometry from
$L^2(\R^D)$ into $\cF_D$.
The Schr\"{o}dinger representation 
$\rho(p, q, \tau)$ of $\sH_D$ on $L^2(\R^D)$ 
can be transferred by
the Bargmann transform to 
a representation $\beta$ of $\sH_D$ on
$\cF_D$.
The \textit{Bargmann--Fock representation} $\beta$ of
$\sH_D$ is defined by
\[
\beta(x, \tau) \sB =\sB \rho(p, q, \tau)
\]
with (\ref{eqn:x_pq}). 
We can verify that (\ref{eqn:Sch_rep1}) is 
mapped to
\[
\beta(y, \tau) F(x) 
=e^{2 \sqrt{-1} \tau -|y|^2/2 - x \cdot \overline{y}}
F(x+y), \quad
y \in \C^D, \quad 
F \in \cF_D.
\]

For $n=(n^{(1)}, \dots, n^{(D)}) \in \N_0^D$ 
and $x=(x^{(1)}, \dots, x^{(D)}) \in \C^D$, we use the notations,
$n! :=\prod_{\ell=1}^D n^{(\ell)} !$ and
$x^n := \prod_{\ell=1}^D (x^{(\ell)})^{n^{(\ell)}}$.
Then a complete orthonormal system (CONS) for $\cF_D$ is given by
\begin{equation}
\varphi_n(x) := \frac{x^n}{\sqrt{n !}}, \quad
n \in \N_0^D, \quad x \in \C^D,
\label{eqn:phi1}
\end{equation}
that is, 
$\bra \varphi_n, \varphi_{m} 
\ket_{L^2(\C^D, \lambda_{\rN(0, 1; \C^D)})}
=\delta_{n m} :=\prod_{\ell=1}^N \delta_{n^{(\ell)} m^{(\ell)}}$,
$n, m \in \N_0^D$. 
Hence, if we define
\begin{align}
k_y(x) &:= \sum_{n \in \N_0^D} \varphi_n(x) \overline{\varphi_n(y)}
=\prod_{\ell=1}^D \sum_{n^{(\ell)} \in \N_0} 
\frac{(x^{(\ell)})^{n^{(\ell)}} (\overline{{y}^{(\ell)}})^{n^{(\ell)}}}{n^{(\ell)} !}
\nonumber\\
&= \prod_{\ell=1}^D 
e^{x^{(\ell)} \overline{y^{(\ell)}}}
=e^{x \cdot \overline{y}},
\label{eqn:reproducingK}
\end{align}
then it works as the \textit{reproducing kernel} of $\cF_D$; 
$F(y)=
\bra F, k_y \ket_{L^2(\C^D, \lambda_{\rN(0, 1; \C^D)})}$
$\forall F \in \cF_D$, $\forall y \in \C^D$. 
This is identified with 
the \textit{correlation kernel} $K_{\sH_D}$ 
of the Heisenberg DPP (\ref{eqn:KHD}) 
given in Definition \ref{thm:HeisenbergDPP}. 

A geometric picture of $\sH_D$ is given in
Chapter XII in 
\cite{Ste93} as follows.
Consider the unit ball in $\C^{D+1}$, 
\[
\B_1^{(2(D+1))}
:=\left\{
w=(w^{(1)}, \dots, w^{(D+1)}) \in \C^{D+1} :
\sum_{\ell=1}^{D+1} |w^{(\ell)}|^2 < 1 \right\}.
\]
By the correspondence
\[
z^{(\ell)}=\frac{2 w^{(\ell)}}{1+w^{(D+1)}}, 
\quad \ell=1, \dots, D, \quad
z^{(D+1)}=\sqrt{-1} \frac{1-w^{(D+1)}}{1+w^{(D+1)} },
\]
$\B_1^{(2(D+1))}$ is mapped to 
an `upper half-space' of $\C^{D+1}$,
\[
\cU^D:=
\left\{ z =(z^{(1)}, \dots, z^{(D+1)}) \in \C^{D+1} :
\Im z^{(D+1)} > \frac{1}{4} \sum_{\ell=1}^D |z^{(\ell)}|^2 \right\}.
\]
Note that the relations between $\B_1^{(2(D+1))}$ and 
$\cU^D$, $D \in \N$ can be regarded as the higher
dimensional extensions of the relation between
the unit disk $\D:=\{w \in \D : |w|<1\} \subset \C$ and
the upper half-plane
$\HH := \{ z : \Im z > 0\}$ via the Cayley transform, 
$z=\sqrt{-1} (1-w)/(1+w)$.
Let $b \cU^D$ be the boundary of $\cU^D$;
$b \cU^D:=
\{ z \in \C^{D+1} :
\Im z^{(D+1)} = \frac{1}{4} \sum_{\ell=1}^D |z^{(\ell)}|^2 \}.
$
We can identity $\sH_D$ with 
$b \cU^D$ by the correspondence,
\[
\sH_D \ni (x, \tau) \quad
\longleftrightarrow \quad
\left(x, \tau + \frac{\sqrt{-1}}{4} |x|^2 \right) \in b \cU^D
\]
with (\ref{eqn:x_pq}). 
In this sense, the reproducing kernel
of the Bargmann--Fock space $\cF_D$
will be interpreted as the Szeg\H{o} kernel
associated with an integral on
the boundary $b \cU^D$ of $\cU^D$ \cite{Zel01,BSZ00} .

%%%%%%%%%%%%%%%%%%%%%%%%%%%%%%%%%%%%%%%%%%%%%%%%%%%%%%%%
%%%%%REFERENCES%%%%%%%%%%%%%%%%%%%%%%%%%%%%%%%%%%%%%%%
%%%%%%%%%%%%%%%%%%%%%%%%%%%%%%%%%%%%%%%%%%%%%%%%%%%%%%%%

%%%%%%%%%%%%%%%%%%%%%%%%%%%%%%%%%%%%%%%%%%%%%%%%%%%%%%%%%%
%%%%%%%%%%%%%%%%%%%%%%%%%%%%%%%%%%%%%%%%%%%%%%%%%%%%%%%%%%
\end{document}